# Does prior knowledge in the form of multiple low-dose PET images (at different dose levels) improve standard-dose PET prediction?

Behnoush Sanaei[1], Reza Faghihi[1], and Hossein Arabi[2]

[1] Division of Medical Radiation, Department of Nuclear Engineering, Shiraz University, Shiraz, Iran

[2] Division of Nuclear Medicine and Molecular Imaging, Department of Medical Imaging, Geneva University Hospital, CH-1211 Geneva 4, Switzerland

[3] Geneva University Neurocenter, Geneva University, 1205 Geneva, Switzerland

[4] Department of Nuclear Medicine and Molecular Imaging, University of Groningen, University Medical Center Groningen, Groningen, Netherlands

[5] Department of Nuclear Medicine, University of Southern Denmark, DK-500, Odense, Denmark

**Running title:** Standard-dose prediction from multiple-low dose PET

**Abstract**

Reducing the injected dose would result in quality degradation and loss of information in PET imaging. To address this issue, deep learning methods have been introduced to predict standard PET images (S-PET) from the corresponding low-dose versions (L-PET). The existing deep learning-based denoising methods solely rely on a single dose level of PET images to predict the S-PET images. In this work, we proposed to exploit the prior knowledge in the form of multiple low-dose levels of PET images (in addition to the target low-dose level) to estimate the S-PET images. To this end, a high-resolution Resnet (HighResNet) architecture was utilized to predict S-PET images from 6% and 4% L-PET images. For the 6% low-dose PET imaging two models were developed, first model was trained using single input of 6% L-PET, and second model was trained using three input channels getting 6%, 4%, and 2% low-dose data as input to predict S-PET images. Similarly, for 4% low-dose PET imaging, a model was trained using single input of 4% low-dose data, and a three-channel model was developed getting 4%, 3%, and 2% L-PET images as input. The performance of the four models was evaluated using structural similarity index (SSI), peak signal-to-noise ratio (PSNR), root mean square error (RMSE) within the entire head regions and malignant lesions. The quantitative analysis of 40 patients demonstrated the superiority of the multi-input network (exploiting the prior knowledge in the form of multiple low dose PET images) over the single-input model. The 4% multi-input model led to improved image quality in terms of SSI and PSNR metrics, and significant decrease in RMSE by 22.22% and 25.42% within the entire head region and malignant lesions, respectively. Furthermore, the 4% multi-input network remarkably decreased the $SUV_{mean}$ bias within the lesions by 64.58% in comparison to single-input network. In addition, the 6% multi-input network led to significant decreases in RMSE within the entire head region, within the lesions, and lesions' $SUV_{mean}$ bias by 37.5%, 39.58%, and 86.99%, respectively. This study demonstrated the significant benefits of using prior knowledge in the form of multiple low-dose PET images to predict S-PET images.

## 1. Introduction

Positron emission tomography (PET) imaging is widely used as an essential tool for many clinical applications such as cancer diagnosis, tumor detection, evaluation of the lesion malignancy, staging of diseases, and treatment follow up [1, 2]. Injection of a standard dose of radioactive tracer is normally required to achieve high-quality PET images in clinical settings. Though injection of high doses of radiopharmaceutical reduces the statistical noise and consequently leads to a better-quality PET images, it raises concerns due to the increased risk of radiation exposure to the patients and healthcare providers [3-5]. On the other hand, using a reduced dose of radioactive tracer would result in quality degradation in PET images due to the increased noise levels and loss of signal (low signal-to-noise-ratio (SNR)) [6-8].

To address this issue, many efforts have been made to improve the quality of low-dose PET (L-PET) images through application of iterative reconstruction algorithms [9-12], image filtering and post-processing [13, 14], and machine learning (ML) methods [4, 15]. In iterative reconstruction algorithms, penalized reconstruction kernels and/or prior knowledge from the anatomical images (MR sequences) are exploited to regularize and/or guide PET image reconstruction to suppress the excessive noise induced by the reduce injected dose [12, 16]. There are two categories of post-reconstruction image denoising including image denoising techniques in the spatial domain and transform domain [13, 17]. These approaches may suffer from some drawbacks such as generating artifacts and/or pseudo signals, over smoothing, hallucinated structures, and high computational time. To address these issues, deep learning (DL) methods, as a special type of machine learning methods, have been dedicatedly developed in which a relationship between L-PET images and standard-dose PET (S-PET) images is learned to predict S-PET images from their low-dose counterparts [7, 18, 19]. Regarding the extraordinary performance of the deep learning methods [18], a number of deep learning-based solutions has been proposed for estimation of the high-quality PET from the low-dose versions with or without aid of anatomical images [20, 21].

Xu et al. [22] employed a U-Net model to predict S-PET images from 0.5% L-PET/MR images. Relying on the generative adversarial networks (GAN) concept, Wang et al. [22] introduced a 3D conditional GAN and Lei et al. [22] proposed a cycle consistent GAN to estimate high quality S-PET images from L-PET images. Besides, Chen et al. [22] implemented a network similar to the model proposed by Xu et al. [22] to predict S-PET from combination of L-PET and MR images and L-PET alone.

The deep learning-based denoising approaches solely rely on a single dose level of L-PET images as the input of models to predict the S-PET images. Given the PET raw data, any low-dose versions of the PET data could be generated. For instance, given a low-dose imaging at 10% of the standard-dose, lower dose levels of the current PET data such as 8%, 6%, 4%, and so on, could easily be generated/reconstructed from the 10% low-dose data. Due to the stochastic nature of PET acquisition, any of these low-dose versions of the PET data would bear complementary/additional information regarding the underlying signal in the standard PET image. In other words, all these lower dose versions of the PET data contain the same or similar signals contaminated with different noise levels and/or distributions. In this light, lower dose PET images could provide prior and/or additional knowledge for prediction of the standard-dose PET images. These prior and/or additional knowledge could be exploited in a deep learning-based denoising framework to enhance the quality of standard dose prediction. To the best of our knowledge, this is the first study employing multiple low-dose levels PET images as prior knowledge to develop a deep learning-based denoising model.

In this study, we investigate the benefits of utilizing additional information in the form of multiple low-dose PET images in a deep learning model. In this regard, we use 6% L-PET imaging data as the input of deep learning model, wherein lower dose PET images with 4% and 2% of standard dose levels (extracted from the raw data of the 6% low-dose PET data) were employed as additional information.

## 2. Materials and Methods

### 2.1. Data acquisition

This study was conducted on PET/CT brain images from 140 patients with head and neck malignant lesions (68 males and 72 females, 71 ± 9 yrs, mean age ± standard deviation (SD)) (100 subjects for training and 40 subjects for evaluation) acquired on a Biograph-6 scanner (Siemens Healthcare, Erlangen, Germany) with standard dose of 210 ± 8 MBq of $^{18}$F-FDG. PET images were acquired for an acquisition time of 20 min, about 40 minutes after the injection. The PET raw data was registered in list-mode format and then 6% low-dose PET data were extracted from the standard data. Then, lower-dose PET data, including 4%, 3%, and 2% were generated from the 6% low-dose data. The low-dose data with the aforementioned percentages were reconstructed using ordered subsets-expectation maximization (OSEM) algorithm with 4 iterations and 18 subsets. The entire PET images were normalized to a range of between 0 to about 1 using a fixed normalization factor for the entire dataset.

**2.2 Deep neural network architecture**

We adopted NiftyNet, an open-source convolutional neural networks (CNNs) platform for deep learning solutions in Python environment, to estimate S-PET images from L-PET images. The S-PET prediction was carried out using a high-resolution Resnet (HighResNet) model. This network consists of nineteen $3 \times 3 \times 3$ convolution layers and one final $1 \times 1 \times 1$ convolution layer. In the first seven layers with 16 kernels, the low-level features of the images such as corners and edges are extracted. The twelve subsequent convolution layers with 32 and 64 kernels, and the final layer with 160 kernels are designed to capture medium and high-level features. Four different models were separately developed to compare/investigate the benefits of employing the multiple low-dose image as the additional information to the denoising network. The deep learning models were assessed to predict the standard-dose PET images from 6% and 4% low-dose data separately. For the 6% low-dose level, first, a model with a single input of 6% was first developed. Then, this model was compared with the model with 3-input channels getting 6%, 4%, and 2% low-dose data as input. Similarly, for 4% low-dose data, first, a model with only a single input for 4% low-dose data was developed. Then, the model was compared with a model with 3-input channels getting 4%, 3%, and 2% as input. The same standard dose PET images were considered as reference for the training of these four models.

**2.3 The details of implementation**

The training of the four models were performed using a L2-norm loss function (the best performance was observed with this loss function) with Adam optimizer, and leaky-ReLU as activation function. The models were developed in 2-dimensional mode with batch-size of 20, weight decay of $10^{-4}$, and max iteration of $10^4$. The training was continued to reach the plateau of the training loss. 5% of the dataset was considered as the evaluation set within the training process. No significant overfitting was observed during the training of the denoising models. The starting learning rate was set at 0.1 which was multiplied by 0.05 every 100 iterations following the recommendation by [23].

**2.4 Evaluation strategy**

To evaluate the prediction performance of the different denoising models and the benefits of applying the prior knowledge in the form of multiple L-PET images, we utilized four quantitative metrics including structural similarity index (SSI), peak signal-to-noise ratio (PSNR), and root mean square error (RMSE) based on the standard uptake value (SUV) within the entire head region and the malignant lesions. Manual segmentation of the lesions, drawn by specialist, in the test dataset were available, thus quantitative metrics were calculated separately for the malignant lesions. These metrics were calculated using following equations:

$$SSI(S,L) = \frac{(2\mu_S\mu_L+c_1)(2\sigma_{SL}+c_2)}{(\mu_S^2+\mu_L^2+c_1)(\sigma_S^2+\sigma_L^2+c_2)} \quad (1)$$

$$c_1 = (0.01d)^2 \qquad c_2 = (0.03d)^2$$

where $\mu_S$, $\mu_L$, and $\sigma_S^2$, $\sigma_L^2$ denote the average and the variance of the S-PET and L-PET images, respectively. $\sigma_{SL}$ represents the covariance of the S-PET and L-PET images, and $d$ is the dynamic range of the pixel values.

$$MSE = \frac{\sum_{M,N}(S(M,N)-L(M,N))^2}{M*N} \quad (2)$$

$$PSNR = 10\log_{10}(\frac{R^2}{MSE}) \quad (3)$$

where $S(M,N)$ and $L(M,N)$ represent the S-PET and L-PET image matrixes with $M$ and $N$ rows and columns. $R$ is the maximum value of the images.

$$RMSE = \sqrt{\frac{1}{N}\sum_N(SUVs_i - SUVl_i)^2} \quad (4)$$

where $SUV_{si}$ and $SUV_{li}$ are SUV at voxel $i$ for the S-PET and L-PET images, respectively. $N$ denotes the total number of image voxels.

In addition to the abovementioned metrics, we calculated the bias of mean SUV ($SUV_{mean}$ bias) on the predicated S-PET images versus the reference S-PET images for the malignant lesions. This metric was calculated using equation 4, wherein $SUV_{mean,S}$ and $SUV_{mean,L}$ are mean standard uptake value within the lesion for the S-PET and L-PET images, respectively.

$$SUV_{mean}\ bias(\%) = 100 \times \frac{SUV_{mean,S}-SUV_{mean,L}}{SUV_{mean,S}} \quad (5)$$

## 3. Results

In order to visually compare the quality of the predicted PET images, sagittal views of the S-PET images predicted by the single-input and the multi-input networks together with the ground truth S-PET image, the corresponding CT image, and the L-PET images of a representative patient are shown in Figures 1 and 2. The increased levels of signal-to-noise-ratio (SNR) were observed when using multi-input models compared to the single-input models.

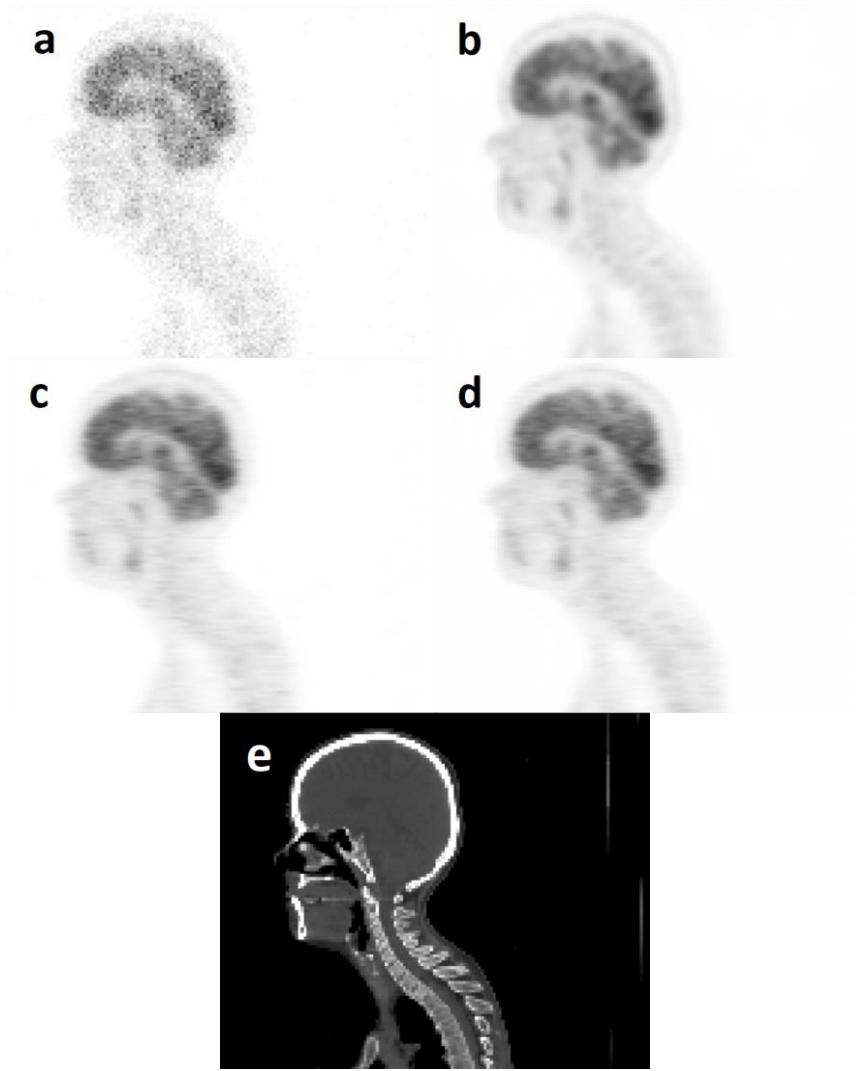

**Figure 1.** Sagittal views of, a) 4% L-PET image, b) ground-truth S-PET image, c) S-PET predicted by the single-input network (4% L-PET image), d) S-PET predicted by the multi-input network (3% and 2% L-PET images as prior knowledge, in addition to 4% L-PET image), and e) CT image.

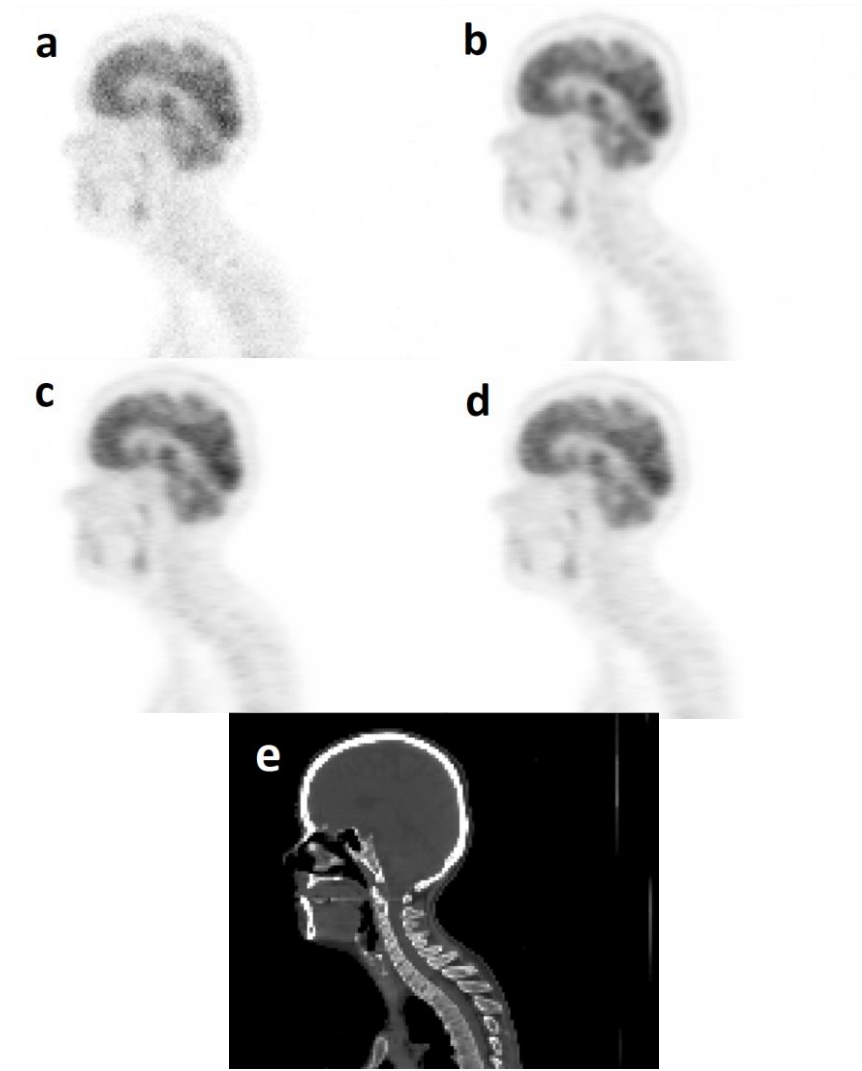

**Figure 2.** Sagittal views of, a) 6% L-PET image, b) ground-truth S-PET image, c) S-PET predicted by the single-input network (6% L-PET image), d) S-PET predicted by the multi-input network (4% and 2% L-PET images as prior knowledge, in addition to 6% L-PET image), and e) CT image.

To quantitatively assess the quality of predicted PET images, SSI, PSNR, RMSE metrics were calculated within the entire head region as well as the malignant lesions. Furthermore, the SUV bias for the lesions were measured on the S-PET images predicted from single-input and multi-input models versus the ground-truth S-PET images. Table 1 summarizes the average and the standard deviation of the abovementioned metrics calculated across 40 patients in the external test dataset for 4% low-dose PET imaging, wherein 3% and 2% low-dose PET data were used as prior knowledge. Besides, Table 2 reports the same metrics calculated for low-dose PET imaging with 6% of S-PET data. The differences between the predicted S-PET images by the single-input and multi-input models were assessed with the paired t-test analysis (considering P-values of smaller than 0.05 statistically significant). The paired t-test analysis demonstrated statistically significant differences between single-input and multi-input models and the benefits of employing the prior knowledge in the form of lower-dose PET images.

**Table 1.** Quantitative metrics calculated for the S-PET images predicted by the single-input (4% L-PET image), and multi-input (4%, 3%, and 2% L-PET images) deep learning models. P-values are calculated between the single-input and multi-input denoising models.

|  | L-PET (4%) | Single-input (4%) | Multi-input (4%, 3%, 2%) | P-value |
|---|---|---|---|---|
| PSNR±SD | 29.59±1.26 | 39.73±2.75 | 41.87±2.81 | 0.01 |
| SSI±SD | 0.86±0.033 | 0.97±0.005 | 0.98±0.003 | 0.04 |
| RMSE±SD (Head) | 0.30±0.04 | 0.09±0.03 | 0.07±0.02 | 0.01 |
| RMSE±SD (Lesion) | 1.19±0.23 | 0.59±0.24 | 0.44±0.18 | <0.01 |
| $SUV_{mean}$ bias ±SD(%) (Lesion) | -0.01±0.93 | -1.92±1.53 | -0.68±0.77 | <0.01 |

**Table 2.** Quantitative metrics calculated for the S-PET images predicted by the single-input (6% L-PET image), and multi-input (6%, 4%, and 2% L-PET images) deep learning models. P-values are calculated between the single-input and multi-input denoising models.

|  | L-PET (6%) | Single-input (6%) | Multi-input (6%, 4%, 2%) | P-value |
|---|---|---|---|---|
| PSNR±SD | 35.33±1.27 | 41.44±2.90 | 44.89±2.89 | 0.01 |
| SSI±SD | 0.93±0.016 | 0.98±0.002 | 0.99±0.001 | 0.04 |
| RMSE±SD (head) | 0.15±0.02 | 0.08±0.02 | 0.05±0.02 | <0.01 |
| RMSE±SD (lesion) | 0.61±0.11 | 0.48±0.20 | 0.29±0.13 | <0.01 |
| $SUV_{mean}$ bias ±SD(%) (Lesion) | 0.09±0.45 | -5.46±0.65 | -0.71±0.49 | <0.01 |

Moreover, to meticulously investigate the performance of the models, the boxplots of RMSE and $SUV_{mean}$ bias within the malignant lesions are presented in Figures 3 and 4.

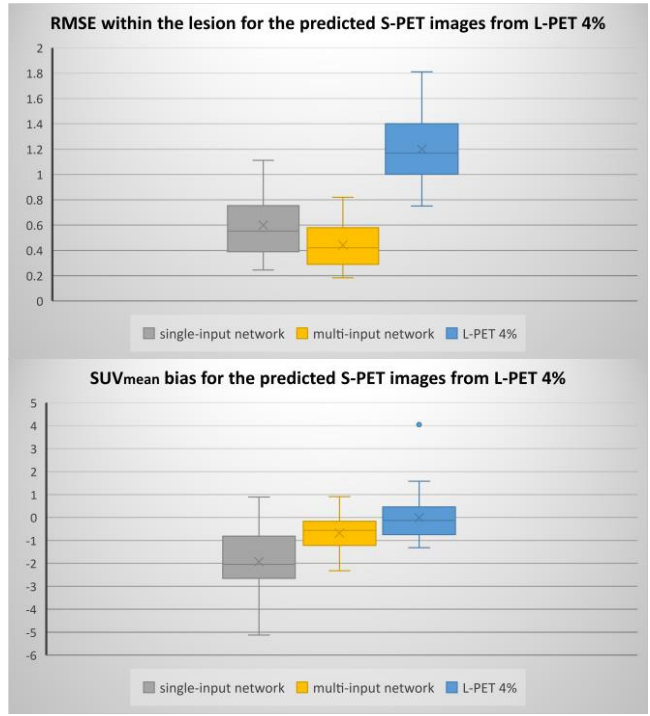

**Figure 3.** Boxplots of SUV$_{mean}$ bias and RMSE within the lesions for the different models and 4% L-PET.

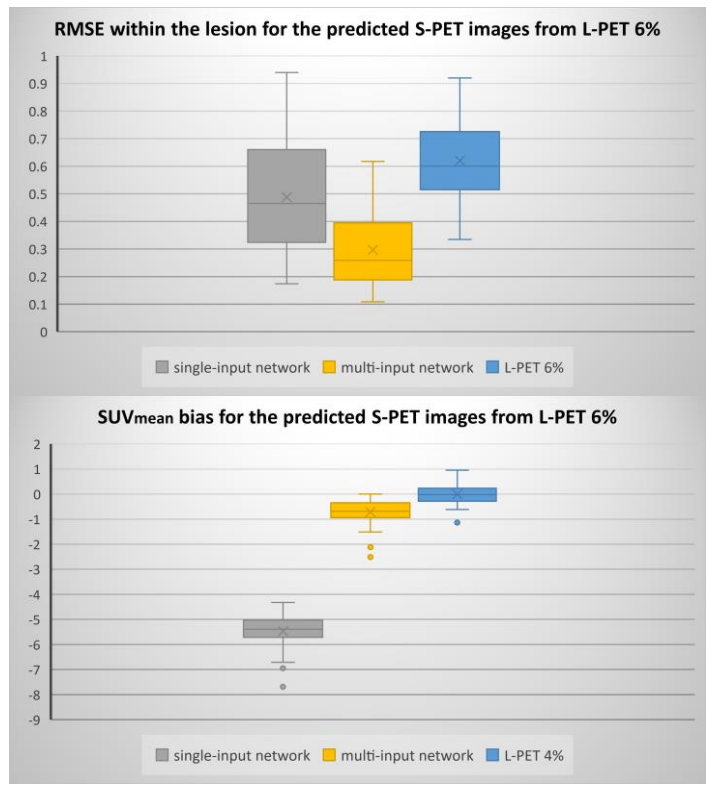

**Figure 4.** Boxplots of SUV$_{mean}$ bias and RMSE within the lesion regions for the different models and 6% L-PET.

## 4. Discussion

In this study, we set out to examine the benefits of employing the prior knowledge in the form of lower-dose PET images for the task of normal PET prediction from low-dose PET data. In this regard, the quality of the predicted S-PET images from the L-PET images at different 4% and 6% dose levels of the standard dose was assessed using two models with single-input and multi-input architectures. As shown in Figures 1 and 2, the quality of predicted PET image improved for both 6% and 4% low-dose imaging when multi-input models were employed getting lower-dose PET images as complementary knowledge. Regarding Figures 1 and 2, the multi-input networks resulted in better image quality compared to single-input networks.

The analysis of the results in Table 1 indicates the superiority of the network trained with multi-input L-PET images over the single-input. The results in Table 1 demonstrated that the 4% multi-input network (using 2% and 3% L-PET as the input in addition to 4% L-PET) would lead to a significant decrease in RMSE by 22.22% (from 0.09 to 0.07) and by 25.42% (from 0.59 to 0.44) within the entire head region and lesions, respectively, compared to the single-input model (using only 4% L-PET images as input). Furthermore, the network trained with multiple L-PET images produces higher values for the SSI and PSNR metrics. It should be emphasized that utilizing prior knowledge in the form of multi-input network remarkably decreases the $SUV_{mean}$ bias within the lesions by 64.58% (from 1.92% to 0.68%).

Regarding Table 2, a similar trend was observed in the analysis of the predicted image by the network trained with 6% L-PET images. SSI and RMSE metrics significantly improved with multi-input network (taking 4% and 2% L-PETs in addition to 6% L-PET). In addition, utilizing multi-input network reduced RMSE within the head region, RMSE within the malignant lesions, and $SUV_{mean}$ bias in the lesions by 37.5%, 39.58%, and 86.99%, respectively.

Figures 3 and 4 compared the distributions of the $SUV_{mean}$ bias and RMSE within the lesions for both single-input and multi-input models. To determine the statistical significance of the differences between these metrics, P-values calculated for $SUV_{mean}$ bias and RMSE (within the lesion) demonstrated significant improvement achieved through employing low-dose prior knowledge in the denoising models.

The primary aim of this work was to investigate the benefits of employing lower-dose PET data in the denoising models for prediction of the standard PET images. Owing to the stochastic nature of PET acquisition and signal formation, reconstruction of the PET data at different low-dose (or low-count) levels would provide different distributions of noise and presentations of signal to noise ratio. Since the underlying signals/uptake patterns in these images are the same, and they are contaminated with different noise levels and/or distributions, they would provide valuable knowledge to the model to distinguish between the noise and underlying signals. The benefits of lower-dose prior knowledge were demonstrated for two low-dose levels of 6% and 4%, wherein significant improvement was observed in both models. The proposed framework could be used in any denoising models where the raw PET (or SPECT) data are available, and the image reconstruction could be performed with different count/dose levels. Moreover, this framework is applicable to the models implemented in the sinogram or projection domain [7, 24-26].

## 5. Conclusion

In this paper, we applied prior knowledge/additional information to the deep learning-based denoising models via utilizing multiple dose levels of L-PET data as the extra input channels to network to estimate the S-PET images. The quantitative evaluation of the proposed framework demonstrated the benefits of employing lower-dose PET data in the denoising models for prediction of the standard PET images. The proposed framework was examined for 6% and 4% low-dose imaging levels. This study recommends using of multiple levels of low-dose PET imaging as prior knowledge to predict the standard-dose PET images.

**Acknowledgments**

??? to be filled by Dr. Faghihi and Prof. Zaidi

**References**

[1] S. Basu, S. Hess, P. E. Nielsen Braad, B. B. Olsen, S. Inglev, and P. F. Høilund-Carlsen, "The Basic Principles of FDG-PET/CT Imaging," (in eng), *PET Clin,* vol. 9, no. 4, pp. 355-70, v, Oct 2014, doi: 10.1016/j.cpet.2014.07.006.

[2] L. Zimmer, "[PET imaging for better understanding of normal and pathological neurotransmission]," (in fre), *Biol Aujourdhui,* vol. 213, no. 3-4, pp. 109-120, 2019, doi: 10.1051/jbio/2019025. L'imagerie TEP pour une meilleure compréhension de la neurotransmission normale et pathologique.

[3] Z. Khoshyari-morad, R. Jahangir, H. Miri-Hakimabad, N. Mohammadi, and H. Arabi, "Monte Carlo-based estimation of patient absorbed dose in 99mTc-DMSA,-MAG3, and-DTPA SPECT imaging using the University of Florida (UF) phantoms," *arXiv preprint arXiv:2103.00619,* 2021.

[4] A. Sanaat, I. Shiri, H. Arabi, I. Mainta, R. Nkoulou, and H. Zaidi, "Deep learning-assisted ultra-fast/low-dose whole-body PET/CT imaging," (in eng), *Eur J Nucl Med Mol Imaging,* vol. 48, no. 8, pp. 2405-2415, Jul 2021, doi: 10.1007/s00259-020-05167-1.

[5] F. H. Fahey, "Dosimetry of Pediatric PET/CT," (in eng), *J Nucl Med,* vol. 50, no. 9, pp. 1483-91, Sep 2009, doi: 10.2967/jnumed.108.054130.

[6] B. Sanaei, R. Faghihi, and H. Arabi, "Quantitative investigation of low-dose PET imaging and post-reconstruction smoothing," *arXiv preprint arXiv:2103.10541,* 2021.

[7] A. Sanaat, H. Arabi, I. Mainta, V. Garibotto, and H. Zaidi, "Projection-space implementation of deep learning-guided low-dose brain PET imaging improves performance over implementation in image-space," *Journal of Nuclear Medicine,* p. jnumed. 119.239327, 2020.

[8] N. Aghakhan Olia *et al.*, "Deep learning-based denoising of low-dose SPECT myocardial perfusion images: quantitative assessment and clinical performance," (in eng), *Eur J Nucl Med Mol Imaging,* Nov 15 2021, doi: 10.1007/s00259-021-05614-7.

[9] J. A. Case, "3D iterative reconstruction can do so much more than reduce dose," (in eng), *Journal of nuclear cardiology : official publication of the American Society of Nuclear Cardiology,* Aug 2 2019, doi: 10.1007/s12350-019-01827-4.

[10] X. Yu, C. Wang, H. Hu, and H. Liu, "Low Dose PET Image Reconstruction with Total Variation Using Alternating Direction Method," (in eng), *PloS one,* vol. 11, no. 12, p. e0166871, 2016, doi: 10.1371/journal.pone.0166871.

[11] N. Zeraatkar *et al.*, "Resolution-recovery-embedded image reconstruction for a high-resolution animal SPECT system," (in eng), *Phys Med,* vol. 30, no. 7, pp. 774-81, Nov 2014, doi: 10.1016/j.ejmp.2014.05.013.

[12] A. Mehranian and A. J. Reader, "Model-Based Deep Learning PET Image Reconstruction Using Forward-Backward Splitting Expectation-Maximization," (in eng), *IEEE Trans Radiat Plasma Med Sci,* vol. 5, no. 1, pp. 54-64, Jun 23 2020, doi: 10.1109/trpms.2020.3004408.

[13] H. Arabi and H. Zaidi, "Improvement of image quality in PET using post-reconstruction hybrid spatial-frequency domain filtering," *Physics in Medicine & Biology,* vol. 63, no. 21, p. 215010, 2018.

[14] H. Arabi and H. Zaidi, "Non-local mean denoising using multiple PET reconstructions," (in eng), *Ann Nucl Med,* vol. 35, no. 2, pp. 176-186, Feb 2021, doi: 10.1007/s12149-020-01550-y.

[15] L. Zhou, J. D. Schaefferkoetter, I. W. K. Tham, G. Huang, and J. Yan, "Supervised learning with cyclegan for low-dose FDG PET image denoising," (in eng), *Med Image Anal,* vol. 65, p. 101770, Oct 2020, doi: 10.1016/j.media.2020.101770.

[16] J. Bland *et al.*, "MR-Guided Kernel EM Reconstruction for Reduced Dose PET Imaging," (in eng), *IEEE Trans Radiat Plasma Med Sci,* vol. 2, no. 3, pp. 235-243, May 2018, doi: 10.1109/trpms.2017.2771490.

[17] H. Arabi and H. Zaidi, "Spatially guided nonlocal mean approach for denoising of PET images," (in eng), *Med Phys,* vol. 47, no. 4, pp. 1656-1669, Apr 2020, doi: 10.1002/mp.14024.

[18] H. Arabi and H. Zaidi, "Applications of artificial intelligence and deep learning in molecular imaging and radiotherapy," *European Journal of Hybrid Imaging,* vol. 4, no. 1, pp. 1-23, 2020.

[19] H. Arabi, A. AkhavanAllaf, A. Sanaat, I. Shiri, and H. Zaidi, "The promise of artificial intelligence and deep learning in PET and SPECT imaging," *Physica Medica,* vol. 83, pp. 122-137, 2021.

[20] K. T. Chen *et al.*, "Ultra-Low-Dose (18)F-Florbetaben Amyloid PET Imaging Using Deep Learning with Multi-Contrast MRI Inputs," (in eng), *Radiology,* vol. 290, no. 3, pp. 649-656, Mar 2019, doi: 10.1148/radiol.2018180940.

[21] H. Liu, J. Wu, W. Lu, J. A. Onofrey, Y. H. Liu, and C. Liu, "Noise reduction with cross-tracer and cross-protocol deep transfer learning for low-dose PET," (in eng), *Physics in medicine and biology,* vol. 65, no. 18, p. 185006, Sep 14 2020, doi: 10.1088/1361-6560/abae08.

[22] R. Fahrig *et al.*, "Design, performance, and applications of a hybrid X-Ray/MR system for interventional guidance.," *Proceedings of the IEEE,* vol. 96, no. 3, pp. 468-480, 2008.

[23] L. N. Smith, "A disciplined approach to neural network hyper-parameters: Part 1--learning rate, batch size, momentum, and weight decay," *arXiv preprint arXiv:1803.09820,* 2018.

[24] H. Arabi and H. Zaidi, "Assessment of deep learning-based PET attenuation correction frameworks in the sinogram domain," (in eng), *Physics in medicine and biology,* vol. 66, no. 14, Jul 7 2021, doi: 10.1088/1361-6560/ac0e79.

[25] N. A. Olia *et al.*, "Deep learning-based noise reduction in low dose SPECT Myocardial Perfusion Imaging: Quantitative assessment and clinical performance," *arXiv preprint arXiv:2103.11974,* 2021.

[26] N. Aghakhan Olia *et al.*, "Deep learning-based low-dose cardiac gated SPECT: in projection or image domain?," in *28th IEEE NUCLEAR SCIENCE SYMPOSIUM AND MEDICAL IMAGING CONFERENCE*, Japan, 2020.